\documentclass[10pt, paper=a4, UKenglish]{article}

\usepackage[style=numeric, sorting=none, giveninits=true, maxbibnames=5, backend=biber]{biblatex}
\def\Title#1{\begin{center} {\Large #1 } \end{center}}
\def\Author#1{\begin{center}{ \sc #1} \end{center}}
\def\Address#1{\begin{center}{ \it #1} \end{center}}

\newcommand\pubblock{\rightline{\begin{tabular}{l} Proceedings of the CTD 2023\\ \pubnumber\\
         \pubdate  \end{tabular}}}

\newenvironment{Abstract}{\begin{quotation} \begin{center} 
             \large ABSTRACT \end{center}\bigskip 
      \begin{center}\begin{large}}{\end{large}\end{center} \end{quotation}}

\newenvironment{Presented}{\begin{quotation} \begin{center} 
             PRESENTED AT\end{center}\bigskip 
      \begin{center}\begin{large}}{\end{large}\end{center} \end{quotation}}

\def\Acknowledgements{\bigskip  \bigskip \begin{center} \begin{large}
      \bf ACKNOWLEDGEMENTS \end{large}\end{center}}

\def\beq{\begin{equation}}
\def\eeq#1{\label{#1}\end{equation}}
\def\eeqn{\end{equation}}

\def\beqa{\begin{eqnarray}}
\def\eeqa#1{\label{#1}\end{eqnarray}}
\def\eeqan{\end{eqnarray}}

\let\bar=\overbar

\def\Dslash{\not{\hbox{\kern-4pt $D$}}}
\def\dslash{\not{\hbox{\kern-2pt $\del$}}}

\def\msb{{\bar{\ssstyle M \kern -1pt S}}}

\usepackage{lineno}

\usepackage{amsmath, amsfonts}

\usepackage{color}

\usepackage[hidelinks]{hyperref}

\usepackage{graphicx}
\usepackage{subcaption}

\usepackage{cleveref}

\usepackage{booktabs}
\usepackage{multirow}
\usepackage{tabularx}

\usepackage{csquotes}
\usepackage{textcomp}

\usepackage{balance}

\newcommand\pubnumber{PROC-CTD2023-07}

\newcommand\pubdate{\today}

\def\affiliation{
\textsuperscript{1} High-Energy Physics department, Radboud University, the Netherlands \\
\textsuperscript{2} National Institute for Subatomic Physics (Nikhef), the Netherlands\\
\textsuperscript{3} SURF, the Netherlands\\
\textsuperscript{4} Instituto de Física Corpuscular, University of Valencia, Spain\\
\textsuperscript{5} IDAL, Department of Electronic Engineering, ETSE-UV, University of Valencia, Spain\\
\textsuperscript{6} Valencian Graduate School and Research Network of AI (ValgrAI), Spain\\
\textsuperscript{7} Computer Architecture for Embedded Systems, University of Twente, The Netherlands}

\newcommand{\conference}{Connecting the Dots Workshop (CTD 2023)\\
October 10-13, 2023}

\usepackage{fancyhdr}
\definecolor{mygrey}{RGB}{105,105,105}
\begin{document}


\large
\begin{titlepage}
\pubblock

\vfill
\Title{Novel Approaches for ML-Assisted Particle Track Reconstruction\\and Hit Clustering}
\vfill

\Author{
\textsc{Uraz Odyurt}\textsuperscript{1,2}, 
\textsc{Nadezhda Dobreva}\textsuperscript{1}, 
\textsc{Zef Wolffs}\textsuperscript{2}, 
\textsc{Yue Zhao}\textsuperscript{3}, 
\textsc{Antonio Ferrer Sánchez}\textsuperscript{5,6}, 
\textsc{Roberto Ruiz de Austri Bazan}\textsuperscript{4}, 
\textsc{José D. Martín-Guerrero}\textsuperscript{5,6}, 
\textsc{Ana-Lucia Varbanescu}\textsuperscript{7}, 
\textsc{Sascha Caron}\textsuperscript{1,2}}
\Address{\affiliation}
\vfill

\begin{Abstract}
Track reconstruction is a vital aspect of High-Energy Physics (HEP) and plays a critical role in major experiments. In this study, we delve into unexplored avenues for particle track reconstruction and hit clustering. Firstly, we enhance the algorithmic design effort by utilising a \emph{simplified simulator (REDVID)} to generate training data that is specifically composed for simplicity. We demonstrate the effectiveness of this data in guiding the development of optimal network architectures. Additionally, we investigate the application of image segmentation networks for this task, exploring their potential for accurate track reconstruction. Moreover, we approach the task from a different perspective by treating it as a hit sequence to track sequence translation problem. Specifically, we explore the utilisation of Transformer architectures for tracking purposes. Our preliminary findings are covered in detail. By considering this novel approach, we aim to uncover new insights and potential advancements in track reconstruction. This research sheds light on previously unexplored methods and provides valuable insights for the field of particle track reconstruction and hit clustering in HEP.
\end{Abstract}

\vfill

\begin{Presented}
\conference
\end{Presented}
\vfill
\end{titlepage}
\def\thefootnote{\fnsymbol{footnote}}
\setcounter{footnote}{0}
%

\normalsize 


\section{Introduction}
\label{sec:introduction}
Scientific use-cases are becoming more and more data-intensive, with every discipline within the domain of natural sciences relying on experimentation involving sensory data collection, in one way or another. Accordingly, data-intensiveness dictates not only application of computational workflows to gain insights, but also the efficient variations of such workflows.

Take the domain of High-Energy Physics (HEP) for instance, especially focusing on collider-based particle physics experiments. Considering the immense scale of the generated data, leading to latency and throughput requirements, it is rather challenging to design compute-efficient algorithms.

For the most part, currently fielded solutions rely on traditional algorithms. In cases that ML models are employed or being researched, the ML model design effort is strictly ad hoc, heavily relying on human expert knowledge. As such, there are no guarantees, assuring the performance and the efficiency of solutions versus potential alternatives. Such shortcomings can be countered through systematic approaches. Enabling a systematic approach towards ML model design will involve search methodologies, which themselves are known to be computationally expensive. We propose that addressing this challenge will be feasible through complexity reduction and methodical simplification of the problem at hand. Simplification in turn allows for efficient evaluation of a variety of solutions/ML model designs.

\paragraph*{Use-case}
Out of numerous available scientific use-cases, we specifically focus on the subatomic particle track reconstruction challenge, a.k.a., \emph{tracking}. The task of tracking involves the reconstruction and tracing of a particle's trajectory, given the sensory data from detectors such as, ALICE~\cite{Collaboration:2008:ALICE}, ATLAS~\cite{Collaboration:2008:ATLAS}, CMS~\cite{Collaboration:2008:CMS}, and LHCb~\cite{Collaboration:2008:LHCb}, installed at the Large Hadron Collider (LHC). Tracking is crucial to the study of HEP collision experiments and the nature of generated particles. It is present as a post-mortem data processing step in the workflows of all major LHC experiments. Relying on traditional and bespoke algorithms, e.g., Kalman filtering, tracking is notoriously compute-intensive at scale. As such, there is a real need for better solutions to be deployed for the upcoming High-Luminosity LHC upgrade.

\paragraph*{Contribution}
This paper covers our preliminary results derived from tracking solutions involving ML model designs. We introduce four main approaches based on the Transformer~\cite{Vaswani:2017:Attention} and the U-Net~\cite{Ronneberger:2015:UNet} model architectures. Each approach has its unique take on tracking. More precisely:
\begin{itemize}
	\item We introduce our iterative approach to tackle complexity and its facilitating role in design and evaluation of ML-assisted solutions.
	\item Starting from the lower levels of the complexity spectrum, we provide the designs and relevant results for four alternative ML models, addressing the tracking challenge.
    \item We demonstrate the use of our flexible and non-physics-accurate simulation framework, REDVID~\cite{Odyurt:2023:RSHE}, as an efficient data set generation tool.
\end{itemize}

Next to this introduction, we provide a brief overview of required background knowledge in \Cref{sec:background}. Our methodology is elaborated in \Cref{sec:methodology}, followed by our ML model designs in \Cref{sec:ml_model_designs}. \Cref{sec:results,sec:conclusion} cover our preliminary results and the concluding remarks, respectively.

\section{Background}
\label{sec:background}
The two fundamental topics we need to be aware of are the role of tracking in HEP and the use of simulations.

\subsection{HEP experiments and tracking}
In this context, by HEP experiments, we refer to accelerator experiments in which, high-energy subatomic particle collisions occur. Such collisions, i.e., \emph{events}, are the result of either proton-proton or ion-ion collisions. The Large Hadron Collider (LHC) is perhaps the most well-known example of such an accelerator.

Events in turn release a plethora of subatomic particles, for which, the behaviour is studied through \emph{tracking} and \emph{calorimetry}. Sophisticated detectors such as ALICE~\cite{Collaboration:2008:ALICE}, ATLAS~\cite{Collaboration:2008:ATLAS}, CMS~\cite{Collaboration:2008:CMS}, and LHCb~\cite{Collaboration:2008:LHCb}, allow us to measure the footprint of individual particles while travelling in space. These tracking detectors are designed to measure the trajectories of charged particles. They consist of layers of sensitive material, such as silicon detectors, generating electrical signals when charged particles pass through them. These signals are the so called \emph{hits} recorded by the detector. These footprints are not continuous, but discrete recordings and limited by detector density.

\subsection{Simulations for HEP}
Any research/design effort with the goal of composing algorithmic solutions, or improving legacy algorithms, requires large amounts of data. These data sets are to be used for extensive testing and validation of the algorithms' expected characteristics, such as correctness, data processing capacity and performance, computational efficiency and power consumption. The same applies, or rather is strictly required, when it comes to solutions involving ML models.

As HEP experiments are not of the kind to be performed on demand, simulations are the next best thing. Simulations for HEP can be used to study the effects of physics phenomena through the generation of data sets for analyses and algorithm design efforts. There are numerous simulations available, predominantly focusing on physics-accuracy and detector specificity. Examples relevant to the ATLAS detector are Geant4~\cite{Agostinelli:2003:GST}, FATRAS~\cite{Edmonds:2008:FATS} and ATLFAST~\cite{Richter-Was:1998:AFSP}.

\section{Methodology}
\label{sec:methodology}
The challenges imposed by a problem that is too complex can be listed as a lower likelihood of finding a solution directly, longer time spent to find a solution, and a much higher likelihood of arriving at an ad hoc solution. In addition, addressing a complex problem directly, seldom accommodates any consideration regarding \emph{secondary objectives}\footnote{a.k.a., non-functional}, e.g., computational performance or resource consumption efficiency.

Our methodology involves the definitions of multiple complexity levels, $C_i$, from simple ($C_1$) to ground truth ($C_n$). Considering that each $C_i$ contains complexity-inducing characteristics, subset-superset relationship is present between consecutive levels, i.e.,
\[
C_1 \subset C_2 \subset \cdots \subset C_n \texttt{.}
\]

\begin{figure}[htbp]
    \centering
    \begin{subfigure}{0.65\textwidth}
    	\includegraphics[width=\linewidth]{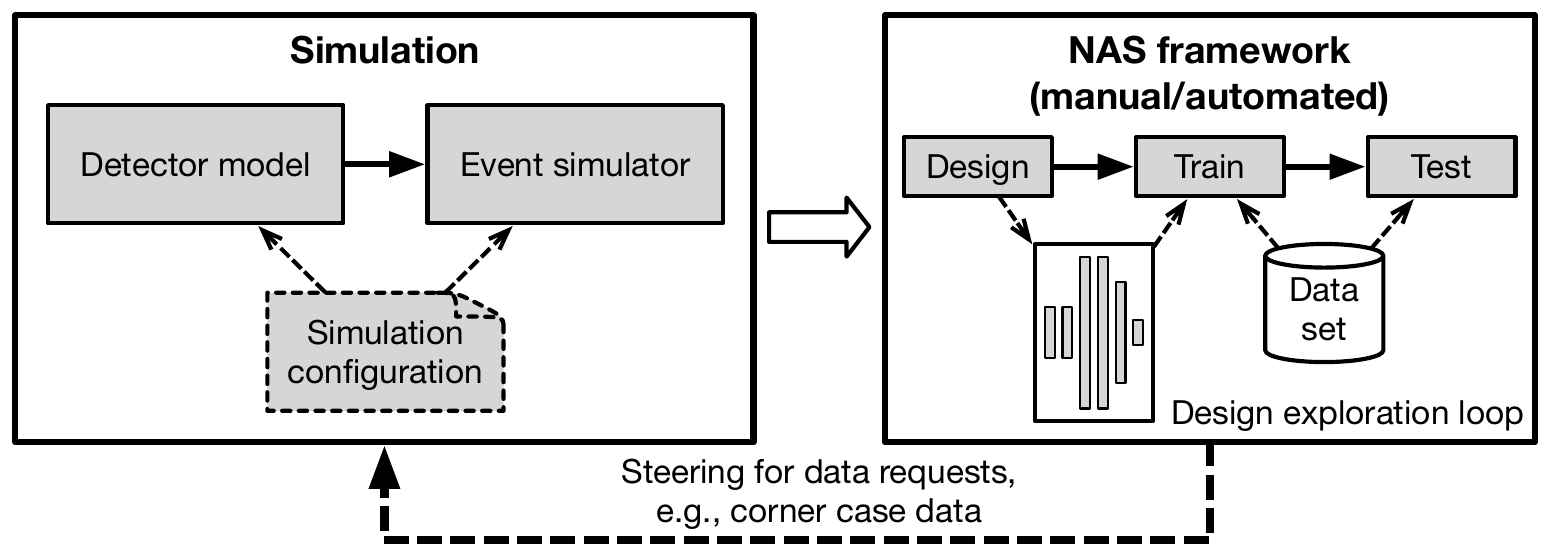}
    	\caption{Complexity level specific data usage}
    	\label{fig:design_search_overview}
    \end{subfigure}
    \qquad
    \begin{subfigure}{0.30\textwidth}
    	\includegraphics[width=\linewidth]{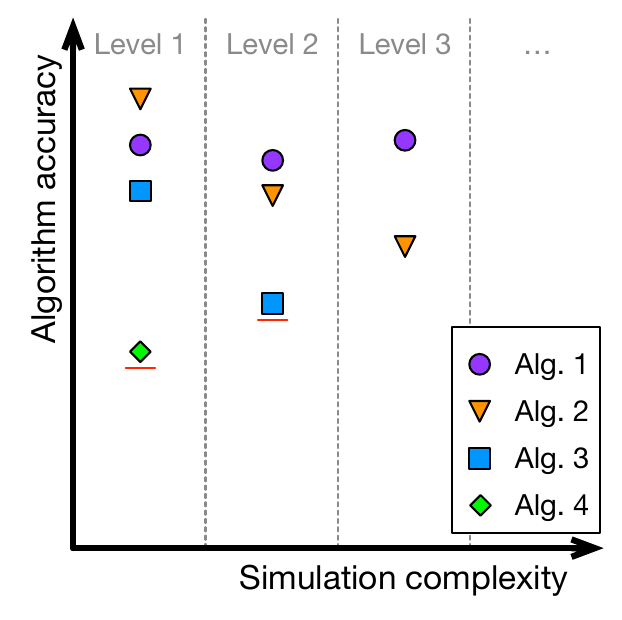}
    	\caption{Solution elimination}
    	\label{fig:design_evaluation_overview}
    \end{subfigure}
    \caption{Methodology framework}
    \label{fig:overview}
\end{figure}

Depending on the difference between levels, a solution addressing a lower level paves the way for a solution addressing a higher level. Our proposed methodology is primarily aimed at automated solution design search. As depicted in \Cref{fig:design_search_overview}, each complexity level will utilise its respective data set, generated with a complexity-aware simulation framework. Our novel simulation framework, REDuced VIrtual Detector (REDVID)~\cite{Odyurt:2023:RSHE}, enables the generation of synthetic data corresponding to different levels of complexity. This is a necessary alternative to the strictly physics-accurate and detector specific simulations discussed in \Cref{sec:background}. With regards to ML models, in this fashion, we are able to efficiently evaluate suitability of designs, starting from the low complexity levels. As higher complexity levels are iteratively incorporated, solution performance can be tracked and inferior solutions can be eliminated \emph{early on}. A conceptual depiction of design evaluation is shown in \Cref{fig:design_evaluation_overview}.

\subsection{Design choices}
Amongst different ML model architectures, we specifically focus on the Transformer and U-Net architectures. A few considerations regarding our deployed architectures are elaborated below.

\paragraph*{Transformer architecture}
As a result of the unordered nature of the recorded hits, we have not considered a positional encoding. For models involving an \emph{encoder-only} design, i.e., EncCla and EncReg as elaborated in \Cref{sec:ml_model_designs}, padding is used to allow variable length inputs. Note that different events could, for instance, have variable track counts, thus resulting in variable numbers of hits.

\paragraph*{U-Net architecture}
Generally speaking, U-Net designs are composed of collections of convolutional layers. Different types of convolutional operators are available, e.g., regular, dilated kernels and submanifold sparse. Our design in its current form incorporates regular convolutions, as the change in performance when using alternatives was negligible.

\subsection{Data sets}
As our data generative tool, REDVID can be configured to apply different levels of detail when simulating HEP events. The most basic level operates within a two-dimensional (2D) geometric space. Though being too simplistic for the purposes of this paper, REDVID simulations in 2D space are pedagogically valuable and are incorporated in student assignments. The results presented in this paper are based on two major levels in three-dimensional (3D) geometric space, with various fine-grained adjustments. Note that the 3D geometric space and elements contained within are defined in cylindrical coordinate system, with $r, \theta$ and $z$ coordinates as radius, angle with the X-axis and location on the Z-axis.

The first level includes randomised linear tracks, resulting in recorded hits. Considering the definition of linear tracks in REDVID, i.e.,
\begin{align*}
r &= r_0 + t \cdot r_d \texttt{,} \\
\theta &= \theta_0 + \theta_d \texttt{,} \\
z &= z_0 + t \cdot z_d \texttt{,}
\end{align*}
the generated data set contains track parameters and the associated hit coordinates per simulated event. Here, $(r_0, \theta_0, z_0)$ is the origin point, while $\langle r_d, \theta_d, z_d \rangle$ is the direction vector. The parameter $t$ is the free variable. The second level includes non-linear track definitions with a helical form, given as
\begin{align*}
r &= r_0 + a \cdot t \texttt{,} \\
\theta &= \theta_0 + d \cdot t \texttt{,} \\
z &= z_0 + b \cdot t \texttt{.}
\end{align*}
Similar to the case of linear tracks, the relevant data set contains track parameters and hit coordinates, though helical track parameters are slightly different and there is no one-to-one match with respect to the case of linear tracks. Similarly, $(r_0, \theta_0, z_0)$ is the origin point, with $a$, $d$ and $b$ representing radial, azimuthal and pitch coefficients, respectively.

Both levels consider hit coordinate smearing and origin smearing. Track counts can be randomised on a per event basis as well.

\section{ML model designs}
\label{sec:ml_model_designs}
We cover four different approaches, based on the Transformer and the U-Net model architectures. Note that the ease of evaluation is enables by the reduction of complexity and simplified simulations.

\subsection{Model 1 - EncDec}
This model closely resembles the original transformer architecture proposed by Vaswani et al.~\cite{Vaswani:2017:Attention}. As such, it has an encoder and a decoder which both make use of a self attention mechanism. The encoder encodes the full set of hits in a given event, and the decoder autoregressively predicts hits belonging to a particular track within the same event.

Of particular interest are the differences with respect to the original transformer architecture. Firstly, this model uses fixed-query attention~\cite{Lee:2019:SetTransformer} in the first encoder stack in order to ensure full positional invariance of the set of input hits. Furthermore, this model also omits positional encoding in the encoder, as the positions of hits are explicitly defined by the coordinates of the hits, which are fed to the model directly. The decoder does use positional encoding on the other hand, as for the constructed track the order of hits is a relevant aspect. The output format is also unique to this model, as rather than predicting a single token with softmaxed probabilities, the decoder outputs a length three vector with $(x, y, z)$ coordinates of the next hit in the track.

\paragraph*{Challenges}
This model differs from the others presented in this work due to the fact that it requires a seed (a short starting sequence of hits) from which to build the track. To do the full reconstruction from hits to tracks it would thus require a preprocessing step to construct track seeds. Furthermore, whereas the other models reconstruct tracks in a one-shot approach, at once creating all tracks in an event, this model builds tracks one by one.

\subsection{Model 2 - EncCla}
This model uses an encoder-only Transformer architecture for the task of classification. It takes sequences of hit features and outputs corresponding sequences of class labels for each hit. The class labels are generated by categorising track parameters into equally-sized bins. These are the track parameters that define the track that each hit belongs to. When there are multiple track parameters, each parameter is put in equally-sized bins, and each unique combination of bins is one class.

The model has 4 Transformer layers, each with 4 attention heads. The classifying layer is a linear layer after the Transformer layers that reduce the data dimensionality from embedding dimensions to the number of classes. Each sequence of data contains all hits in an event. Each batch contains 128 sequences, padded to the length of the longest sequence. The test data set constitutes 20\% of the data. Within the rest of data, there is a 75/25 split between training and validation data sets.

\paragraph*{Challenges}
The main challenge for this approach is the reliance on parameter granularity reduction through binning. As the scale goes up, i.e., the number of tracks per event increases, the number of bins has to increase as well. The ideal case is to have no more than one track in any bin. Note that the distribution of tracks and therefore track parameter values, is not uniform.

\subsection{Model 3 - EncReg}
The third model under consideration is another encoder-only Transformer. It is a sequence-to-sequence model the input of which is the cylindrical hit coordinates of a single event and the output is their corresponding regressed track parameters. The model has an input layer creating an initial 32-dimensional embedding of the coordinates, followed by 6 encoder blocks, a dropout layer and an output layer. Each encoder block has 8 attention heads in its multi-head attention layers and its linear layers have 256 dimensions. The loss function of this model is Mean Squared Loss. This Transformer uses padding up to a maximum number of hits to handle the variable length inputs; the input of the network is this a multi-dimensional tensor with size (batches, max number of hits, data dimensionality). The output is then of size (batches, max number of hits, number of track parameters). It is important to note that different data sets (linear vs helical) require different models, due to the difference in regressed parameters describing the tracks.

To obtain the hit classification, a clustering algorithm is run on the regressed track parameter space. We make use of Agglomerative clustering with a distance threshold optimized for the specific data set. A not yet implemented third stage would consist of a regressor neural network, which takes as input a cluster of hits, and regresses a single track parameter tuple defining the track they belong to.

\paragraph*{Challenges}
Perhaps the biggest challenge for this model is the discovery of track parameters that sufficiently define a track and can be learned by the model. What coordinate system they should be in, dealing with angle symmetry, different weighting of the tracks' contribution to the loss, etc. are some examples of things to consider. Another challenge is the evaluation of the EncReg model: as accuracy cannot be calculated for the regressed values, its performance is indirectly evaluated based on the formed clusters in the stage following it.

\subsection{Model 4 - U-Net}
An alternative methodology under consideration for this task involves the utilisation of a U-Net architecture for the purpose of pixel segmentation, specifically targeting those pixels that correspond to a particle track within preprocessed data. Segmentation is based on the spatial coordinates $(x, y, z)$. The input of the network input comprises a multi-dimensional tensor with size $(n_{\text{batches}}, 1, \text{width}, \text{height}, \text{depth})$, which coincides with the dimensions of the output tensor. The last provides the probability of membership to the ``track'' class for each pixel, with this information encoded in an accessible channel. Subsequently, the classification of each track is derived through a post-processing step, employing density-based clustering methods, such as DBSCAN~\cite{Ester:1996:Density}, after the optimisation of relevant hyperparameters.

The network receives an input tensor of dimensions $(samples, 1, \text{width}, \text{height}, \text{depth})$, encompassing the original data points. The output tensor, maintaining identical dimensions, generates a mask representing the interpolated values between the original dots.

\paragraph*{Challenges}
Arguably, the most significant challenge faced by the U-Net lies in efficiently mastering binary segmentation while accounting for all pixels corresponding to the background, given that these constitute the vast majority ($>$90\%) of the total pixels. At this point, this issue has been addressed by employing a weight three orders of magnitude higher for pixels representing the hits in the Binary Cross Entropy utilised as the loss function.

\subsection{Training notes}
As expected, different model designs are trained best on different combinations of learning hyperparameters. Not every model design covered in this paper is enhanced with hyperparameter optimisation, yet. The current sets of learning parameters per model are given in \Cref{tab:hyperparameters}, with EncDec's training taking advantage of this type of optimisation. Examples of training-specific learning parameters are the Learning Rate (LR) and the batch size, i.e., the batch of training samples to consider before a model update.
\begin{table}[htbp]
    \centering
    \begin{tabular}{@{}lcccccc@{}}
        \toprule
        \multicolumn{1}{c}{Model} & LR      & Batch size & \begin{tabular}[c]{@{}c@{}}Optimisation\\ algorithm\end{tabular} & \begin{tabular}[c]{@{}c@{}}Max\\ \#parameters\end{tabular} & Framework  & Specific info. \\ 
        \midrule
        EncDec                   & 0.0001       & 64         & Adam      & 7127171        & TensorFlow      & \begin{tabular}[c]{@{}c@{}}Weights and Biases~\cite{WB:2023}\\ HP optimisation\end{tabular} \\
        EncCla                   & 0.001        & 128        & AdamW     & 557522         & PyTorch         & PyTorch CrossEntropyLoss \\
        EncReg                   & 0.001        & 16         & Adam      & 126371         & PyTorch         & Agglomerative clustering \\
        U-Net                    & 0.00005      & 10         & Adam      & 6952259        & PyTorch         & \begin{tabular}[c]{@{}c@{}}DBSCAN\\ $eps = 1.5$, $min\_samples = 1$\end{tabular} \\ 
        \bottomrule
    \end{tabular}
    \caption{Summary of the learning parameters considered for the training of our model designs. Note that the \emph{Max \#parameters} refers to the largest model if multiple variations were considered.}
    \label{tab:hyperparameters}
\end{table}

\subsection{Order of tasks}
\begin{figure}[htbp]
    \centering
    \begin{subfigure}{0.8\textwidth}
    	\includegraphics[width=\linewidth]{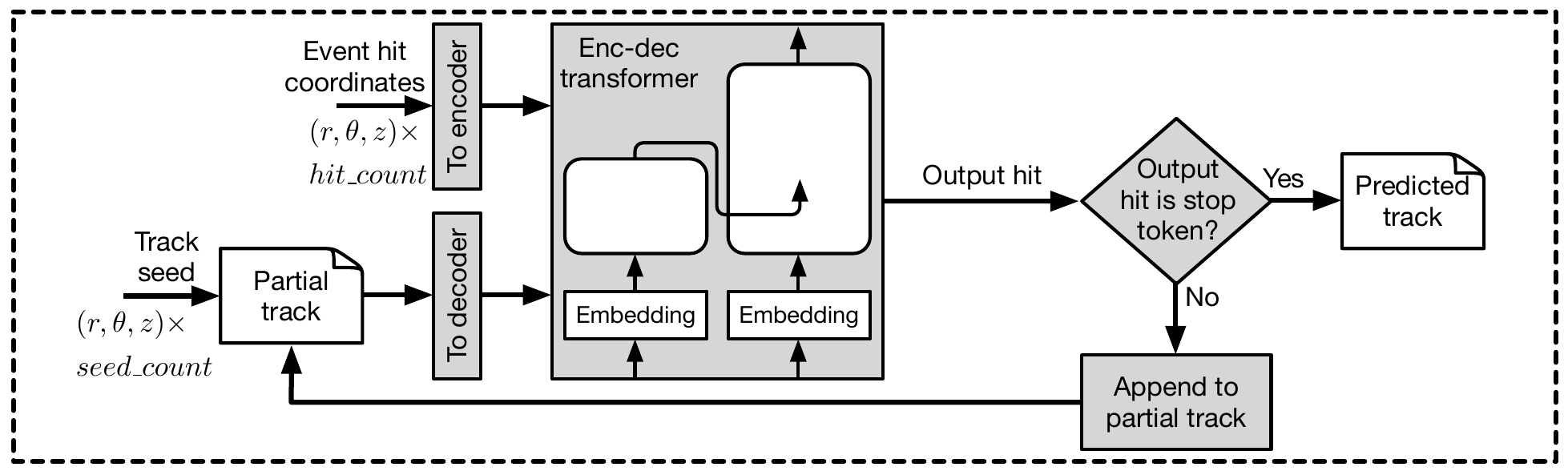}
    	\caption{Model 1 uses a full transformer architecture to autoregressively append hits to a track candidate, starting with a seed. The autoregressive iterations stop whenever the model has output the stop token.}
    	\label{fig:flow_model_1}
    \end{subfigure}
    \qquad
    \begin{subfigure}{0.8\textwidth}
    	\includegraphics[width=\linewidth]{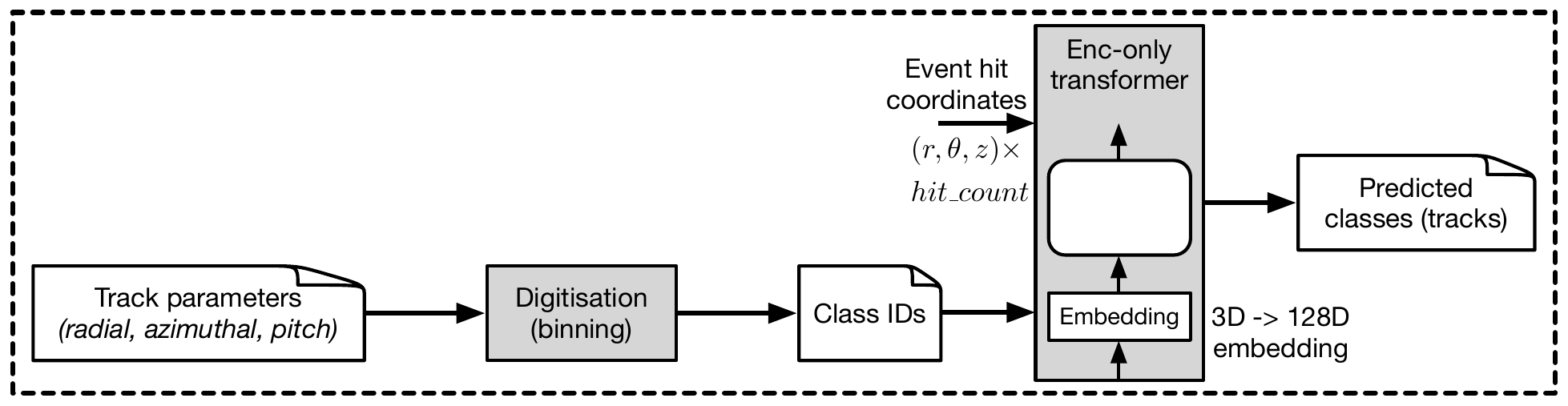}
    	\caption{Model 2, utilising an encoder-only transformer and considering a digitisation step to assign class IDs to tracks. The model assigns class IDs, i.e., bins, to each of an event's hits.}
    	\label{fig:flow_model_2}
    \end{subfigure}
    \qquad
    \begin{subfigure}{0.8\textwidth}
    	\includegraphics[width=\linewidth]{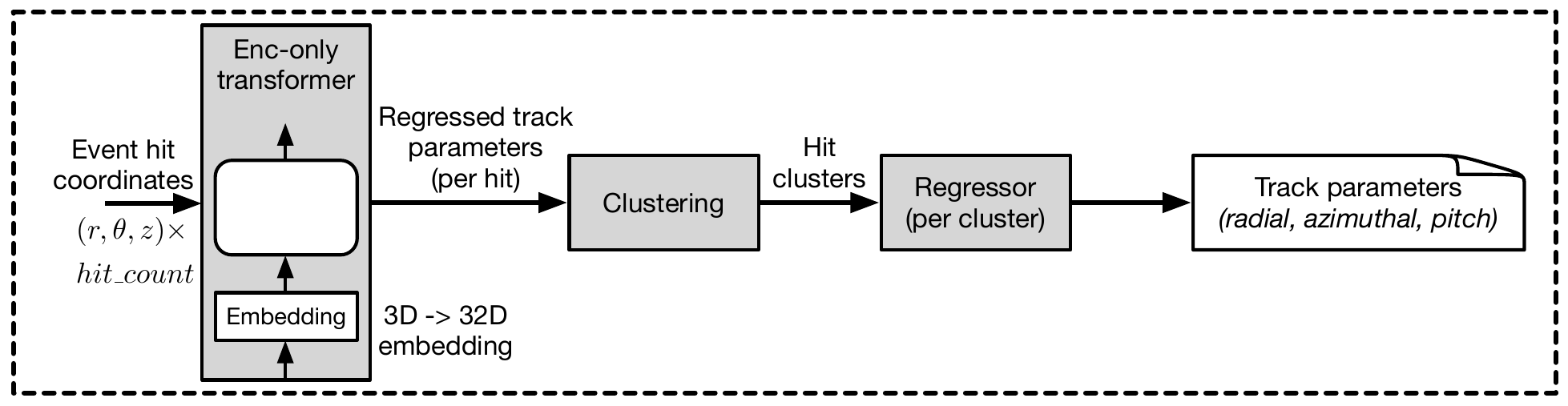}
    	\caption{Model 3, utilising an encoder-only transformer, regressing the track-defining parameters of an event's hits, followed by clustering in the track parameter space, and finally, by per-cluster regression of the track parameters of every particle.}
    	\label{fig:flow_model_3}
    \end{subfigure}
    \qquad
    \begin{subfigure}{0.8\textwidth}
    	\includegraphics[width=\linewidth]{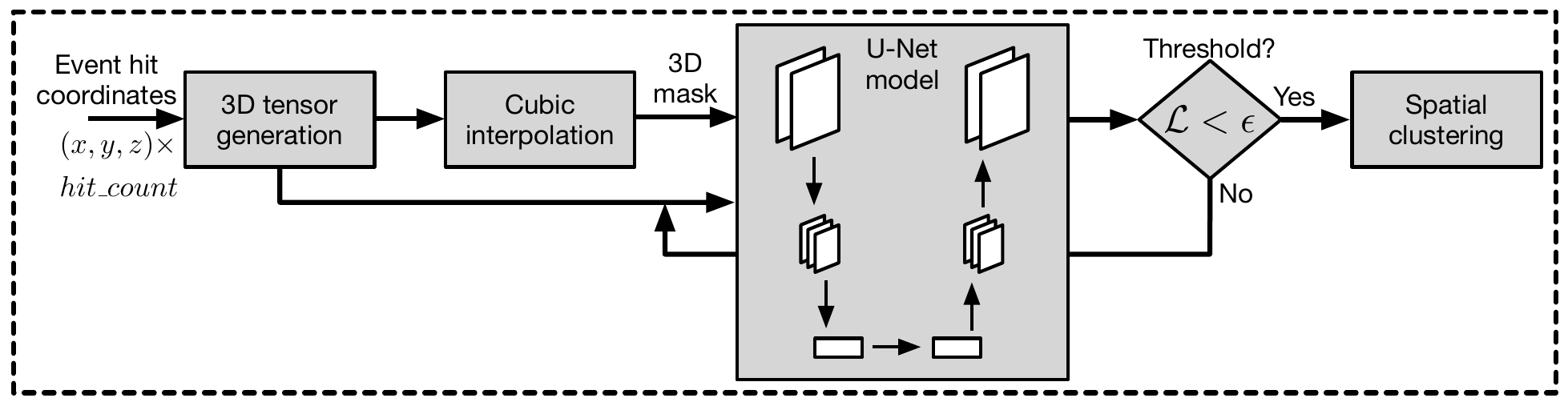}
    	\caption{Model 4 in a U-Net-style approach, with a 3D mask constructed from input hit coordinates in order to train the network until a specific loss threshold is met, or the considered number of iterations is reached. The workflow culminates in a spatial clustering step to label the hits.}
    	\label{fig:flow_model_4}
    \end{subfigure}
    \caption{Depicting the high-level views of dedicated workflows for each ML model design.}
    \label{fig:flow_diagrams}
\end{figure}

Generally, we emphasise the \enquote{ML-assisted} nature of solutions. Oftentimes, aiming for a monolithic ML model as the solution of the challenge at hand, does not yield the best result, if at all. And if it does, the solution will not be a modular or flexible one. As such, our efforts have been channelled into the composition of solution workflows involving ML models, but also other complementary steps, e.g., various pre- or post-processing steps. The high-level view of our workflows, involving our model designs, are depicted in \Cref{fig:flow_diagrams}.

\section{Results and performance}
\label{sec:results}
We have collected the preliminary results of our model training with a variety of data sets, generated using REDVID. The data sets we are reporting preliminary results for are,
\begin{itemize}
    \item 3 helical tracks per event,
    \item 1-20 (variable count) helical tracks per event,
    \item 10-50 (variable count) helical tracks per event,
\end{itemize}

The two model performance metrics we have considered at this point are \emph{prediction accuracy} and the \emph{TrackML score}~\cite{Kiehn:2019:TrackML}, which is a custom metric. In the definition of TrackML, reconstructed tracks with three or more hits are considered, while at least 50\% of a reconstructed track's hits must originate from the same truth particle. Since our simulations do not generate particles, but tracks, as part of the true data, we have replaced the true particle with the true track. The TrackML score considers the score of a track as being the sum of correctly assigned hit weights. For our data set, we consider the weight value 1 for all hits. Available scoring for each model is provided in \Cref{tab:scores}.
\begin{table}[htbp]
    \centering
    \begin{tabular}{@{}lcccccc@{}}
        \toprule
        \multicolumn{1}{c}{Approach} & \begin{tabular}[c]{@{}c@{}}10-50 helical\\ Accuracy; TrackML score\end{tabular} & \begin{tabular}[c]{@{}c@{}}1-20 helical\\ Accuracy; TrackML score\end{tabular} & \begin{tabular}[c]{@{}c@{}}3 helical\\ Accuracy; TrackML score\end{tabular} \\ 
        \midrule
        EncDec                & 85\%; n/a               & 92\%; n/a           & n/a; n/a \\ 
        EncCla                & 88\%; 98\%              & 88\%; 98\%          & 89\%; 98\% \\ 
        EncReg                & n/a; 87\%               & n/a; 94\%           & n/a; 99\% \\ 
        U-Net                 & n/a; n/a                & 97.7\%; n/a         & 99.7\%; n/a \\ 
        \bottomrule
    \end{tabular}
    \caption{Available scores for different models are given per data set, focusing on helical data. Note that data sets can contain different fixed, or variable (randomised) counts of linear/helical tracks per event.}
    \label{tab:scores}
\end{table}

\section{Conclusion and future work}
\label{sec:conclusion}
We have showcased our initial results from four individual ML model designs, based on two architectures, i.e., Transformers and U-Nets. We have discussed the pros and cons of each solution, as well as complementary steps required. On top of that, we have introduced our iterative approach towards complexity, as a methodology in the making. In our opinion, this methodology is an improving step towards systematic ML model design and to make model design a swift process. Last but not least, we elaborated the role of complexity-reduced simulations as an integral part of our methodology.

\Acknowledgements
This paper is a result of the collaboration between Radboud University, Nikhef, the University of Twente, SURF, the University of Valencia and the University of Amsterdam. The project has been supported by grants and computing resources from NWO (grant no.\ 62004546), SURF (grant no.\ EINF-7404), Valencian Community (grant no.\ CIAICO/2021/184), Valencian Foundation (ValgrAI), Quantum Spain project and Artemisa.


\printbibliography

\end{document}